# Room temperature ferromagnetism induced by high valence cation $V^{+5}/V^{+4}$ substitution in $SrFeO_{3-\delta}$

Rakhi Saha[1], Koyal Suman Samantaray[1], P Maneesha[1], SC Baral[1], Sachin Sarangi[2], Rajashri Urkude[3], Biplab Ghosh[3], Abdelkrim Mekki[4,5], Khalil Harrabi[5,6], Somaditya Sen[1]*

[1]*Department of Physics, Indian Institute of Technology Indore, Indore, 453552, India*
[2]*Institute of Physics, P.O. Sainik School, Sachivalaya Marg, Bhubaneswar, 751005, India*
[3]*Beamline Development & Application Section, Bhabha Atomic Research Centre, Trombay, Mumbai 400085*
[4] *Department of Physics, King Fahd University of Petroleum & Minerals Dhahran, 31261, Saudi Arabia*
[5]*Interdiciplinary Research Center (IRC) for Advanced Material, King Fahd University of Petroleum & Minerals, Dhahran 31261, Saudi Arabia*
[6]*Interdiciplinary Research Center (RC) for Intelligent Secure Systems, King Fahd University of Petroleum & Minerals, Dhahran 31261, Saudi Arabia*

*Corresponding author: sens@iiti.ac.in*

## Abstract:

The structural and magnetic effects of non-magnetic vanadium (V) doping in helimagnetic $SrFeO_{3-\delta}$ (SFO) are investigated, focusing on up to 3% substitution at the Fe site. Structural analysis from X-ray diffraction (XRD) and Raman spectroscopy, supported by phonon mode calculations, reveals that pure SFO exists as a mixed tetragonal-orthorhombic phase, while V-doped samples exhibit an emerging cubic phase alongside tetragonal symmetry. Magnetic hysteresis (M-H) loops show notable ferromagnetic behavior within the antiferromagnetic matrix, persisting even at room temperature. Temperature-dependent magnetization measurements indicate a Néel temperature ($T_N$) shift from 70K to 55K, along with increased magnetization differences in field-cooled (FC) and zero-field-cooled (ZFC) data, reflecting heightened magnetic frustration due to competing FM/AFM exchange interactions. X-ray photoelectron spectroscopy (XPS) and X-ray absorption near-edge structure (XANES) analyses reveal a rise in $Fe^{3+}$ and $V^{5+}$ states, affecting oxygen vacancy distributions and corresponding structural shifts seen in XRD and Raman results. The multivalent $Fe^{3+}/Fe^{4+}$ and $V^{4+}/V^{5+}$ states enhance double-exchange (DE) and super-exchange (SE) interactions ($Fe^{3+}$-O-$Fe^{4+}$ and $Fe^{3+}$-O-$V^{5+}$), promoting ferromagnetism. Frequency-dependent magnetization studies display a subtle susceptibility peak shift, indicating spin-glass-like behavior in V-doped samples.

**Introduction:**

Perovskite oxides have adaptable ABO₃ crystal structures. These have been a focal point in materials science research for a long time due to their remarkable electronic, magnetic, and multiferroic properties [1], [2]. One of the most intriguing features is their ability to exhibit a wide range of magnetic and ferroelectric behaviors along with the coexistence of both properties leading to multiferroicity and in some cases magnetoelectric coupling. These diverse properties allow them to be used in cutting-edge applications like spintronics, sensors, and multifunctional devices. Amongst several perovskite ABO₃ oxides, Strontium Ferrite (SrFeO₃, hereafter referred to as SFO) is one of the most interesting materials due to its versatile electronic, magnetic, and catalytic properties correlated to its versatile structural changes associated with O-content. With the loss of O from the lattice various structural phase transitions happen from cubic to tetragonal and further to orthorhombic. The electronic properties get modified due to these structural changes [3]. In SrFeO₃, Fe stays in an unusually high oxidation state 4+. It also shows a helical spin order which is due to competing effect of anti-ferro and ferro magnetic interaction along with metal-like conductivity. Moreover, it shows no John-teller distortion. A report by Lebon et al showed due to the charge ordering of $Fe^{3+}$-$Fe^{4+}$ a large negative magnetic resistance could be observed and this is associated with a metal-to-insulator transition at δ=0.15 with magnetic transition at ~ 70K [4]. Additionally, DFT studies that showed defects due to oxygen vacancies ($V_O$) distort the FeO₆ octahedral volume, which significantly influences the super-exchange interactions within the Fe-O-Fe bonds [5].

In literature, a considerable amount of work has been done on SFO by modifying the oxygen defect state by doping and thereby affecting its magnetic ground state. The d-orbital configurations of Fe and the dopant are susceptible and important in determining the magnetic ground state of the material. Seki et al. [6] showed that a 40% substitution of Fe by Ni induces ferromagnetism above room temperature. A subsequent theoretical analysis attributed this to strong magnetic Ni-O-Fe exchange interactions mediated by a 2p-3d hybridization of O with an unusual high-spin $Fe^{5+}$ ($t^3_{2g}$) and $Ni^{5+}$ ($t^3_{2g}$ $e^2_g$) delocalization [7]. Kinoshita et.al showed that Rh and Ru substitution in SrFeO₃ produced two contrasting outcomes: Rh doping resulted in a ferromagnetic (FM) state, while Ru doping led to a spin-glass state, despite SrRhO₃ being paramagnetic and SrRuO₃ being ferromagnetic [8]. Hence, the individual dopant is not a determinant of the type of magnetism; the combined exchange integral of the dopant with Fe in that particular compound is the determining factor.

In this study, the structural and magnetic property variations of SrFe$_{(1-x)}$V$_x$O$_{3-\delta}$ (0 < x < 0.03) have been studied. Several reports on transition metal (TM) doping at the B-site induced exotic magnetic phases. However, magnetic properties have not been critically explored for V-doped SFO. The mixed high-valence states of $V^{4+}/V^{5+}$, along with $Fe^{3+}/Fe^{4+}$, provided a platform for potential alterations in the helical spin order observed in SFO. Higher oxidation states of V are likely to impact $V_O$, influencing the symmetry of the lattice structure as well. Here we have structurally correlated the effect of incorporating the non-magnetic V ions in the magnetic ground state of the material.

**Experimental methods:**

Nanocrystalline SrFe$_{(1-x)}$V$_x$O$_{3-\delta}$ (0 ≤ x ≤ 0.03) was synthesized using a modified Pechini sol-gel method. Strontium nitrate (99%, Alfa Aesar), iron nitrate nonahydrate (98%, Alfa Aesar), and vanadium(V) pentoxide (99.6%, Alfa Aesar) were utilized as precursors. After dissolving the precursors in solvents, the appropriate proportions of cations and fuel were added to the solution, which was dried overnight at 85°C. The resulting dried black, fluffy powders were ground using a mortar and pestle, and the final samples were calcined at 900°C for 3 hours. These samples were labeled as S0 (x = 0), S1 (x = 0.01), S2 (x = 0.02), and S3 (x = 0.03). The phase purity of the powders was verified by X-ray diffraction (XRD) using a Bruker D2 Phaser diffractometer with Cu Kα radiation (λ = 1.54 Å) over a 2θ range of 20° to 80°, with a step size of 0.02°. Lattice parameters, bond angles, and other structural details were obtained via Rietveld refinement using the Fullprof suite. Room temperature Raman spectra were collected using a HORIBA LabRAM HR 300 Evolution system, covering the range of 30-400 cm$^{-1}$, with a He–Ne laser of wavelength 632.8 nm. For X-ray photoelectron spectroscopy (XPS) measurements, monochromatic Al Kα X-rays (hν = 1486.6 eV) operating at 150W under ultra-high vacuum (~10$^{-9}$ mbar) were employed. A survey spectrum was initially obtained for each sample over a binding energy range of 0 to 1200 eV, followed by high-resolution spectra for Fe2p, Sr3d, V2p, and O1s. The spectra were analysed with the help of XPSPEAK4.1 software. X-ray absorption spectroscopy (XAS) measurements were performed at the Fe-K and V-K edges using beamline-09 at RRCAT to investigate the coordination chemistry. The standard normalization and background subtraction procedures were executed using the ATHENA software version 0.9.26 to obtain normalized XANES spectra. Magnetic measurements, temperature-dependent and frequency dependent magnetization, were carried out in Superconducting Quantum Interference Device (SQUID) using vibrating sample magnetometer (VSM). AC magnetic measurements were also performed at various AC amplitudes with a 200 Oe bias field. The measurements were conducted during heating after cooling the system in zero magnetic field.

**Result and discussion:**

Crystal structure:

The room temperature XRD patterns of the S0 sample [Fig: 1] confirm a perovskite structure, primarily tetragonal (T) I4/mmm (JCPDS PDF no. 40-0906) with a minor orthorhombic (Or) Cmmm (0.5%) phase [9]. Upon V incorporation into the $SrFeO_3$ lattice, the Or phase disappears, and approximately 12% of a cubic (C) Pm-3m phase emerges, coexisting with the T phase. No impurity phases are detected, indicating that V does not introduce other oxide forms. Notably, V stabilizes the cubic structure, as seen from the merging of cleaved peak intensities, reflecting increased lattice symmetry at $2\theta \approx 47°$ [Fig: 1 (inset)]. A slight peak shift toward lower angles suggests a small increase in lattice parameters.

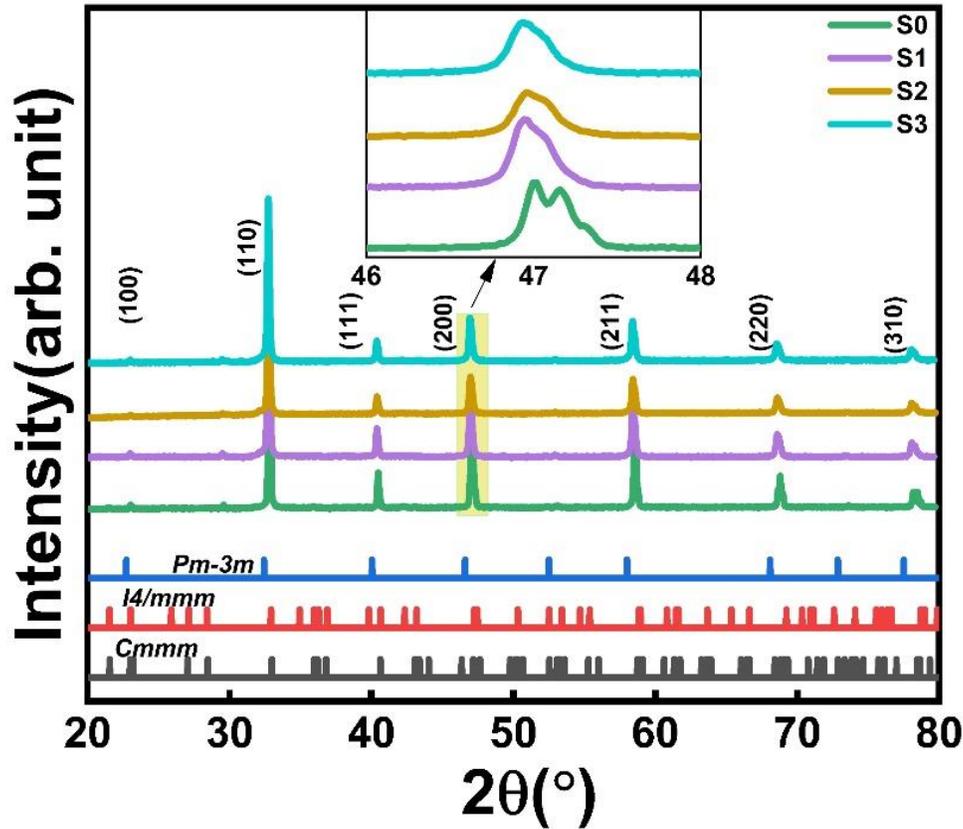

**Fig 1:** X-ray diffraction pattern of all sample with cif file of Pm-3m, I4/mmm and Cmmm space group

Rietveld refinement was performed with mixed tetragonal (T) and orthorhombic (Or) phases for S0, while with mixed T and C phases for the rest of the samples. A good fit was achieved with low R factors and $\chi^2$ values (Table-1). In the T structure of $SrFeO_3$, the Fe site (B-site) exhibits three different crystal fields: a) one $Fe^{(1)}O^{(1)}O^{(2)}_4$ square pyramidal unit, b) two $Fe^{(2)}O^{(3)}_2O^{(2)}_2O^{(4)}_2$ octahedral units, and c) one $Fe^{(3)}O^{(5)}_2O^{(4)}_4$ octahedral units. In contrast, the C structure consists of similar $FeO_6$ octahedra forming long chains. In the T structure, there is a loss of O-atom. Out of the four similar octahedra in the C phase, one octahedral chain loses O, forming the $Fe^{(1)}O^{(1)}O^{(2)}_4$ square pyramids, with a square $O^{(2)}_4$ base and an apical $O^{(1)}$. This $O^{(1)}$ is connected to the next octahedra which has also lost one O-atom and formed a square-planar pyramid and thereby forms a dimer connected through the apical $O^{(1)}$. The dimers have four equal $Fe^{(1)}$-$O^{(2)}$ bonds and a nominally larger $Fe^{(1)}$-$O^{(1)}$ bond with the apical $O^{(1)}$. The crystal field, therefore is no longer an octahedral type but a square-pyramidal type. These dimers are connected through the planar $O^{(2)}$ atom to the neighboring $Fe^{(2)}O^{(3)}{}_2O^{(2)}{}_2O^{(4)}{}_2$ octahedra. Note that these octahedra have three different Fe-O bonds: two planar $Fe^{(2)}$-$O^{(4)}$, two planar $Fe^{(2)}$-$O^{(2)}$, and two apical $Fe^{(2)}$-$O^{(3)}$ bonds. Hence, the $O^{(3)}{}_2$-$O^{(2)}{}_2$-$O^{(4)}{}_2$ octahedral unit has a kite shaped planar $O^{(2)}{}_2$-$O^{(4)}{}_2$ rather than a square. This kite-shaped $O_4$ structure comes from the effect of the crystal field modification of the $Fe^{(1)}O_5$ unit due to the O-loss, thereby modifying the neighboring $FeO_6$ octahedra. The two $Fe^{(2)}O^{(3)}{}_2O^{(2)}{}_2O^{(4)}{}_2$ octahedral units are the immediate neighbors of the dimer units. On the other hand, the $Fe^{(3)}O^{(5)}{}_2O^{(4)}{}_4$ octahedral unit is diagonally opposite to the dimer unit and is composed of four planar $O^{(4)}{}_4$ and two apical $O^{(5)}$ thereby forming four equal $Fe^{(3)}$-$O^{(4)}$ planar bonds and two $Fe^{(3)}$-$O^{(5)}$ apical bonds. This octahedra has a square plane and is symmetric in nature. This four-unit analysis enables one to understand the effect of O-loss in a C structure which leads to a phase transformation to a T structure. Note that for further O-loss the $Fe^{(3)}O^{(5)}{}_2O^{(4)}{}_4$ octahedra gets affected by forming dimers similar to the $Fe^{(1)}O^{(1)}O^{(2)}{}_4$ dimers. In such a situation, the diagonally opposite dimer chains have intermediate neighboring $Fe^{(2)}O^{(3)}{}_2O^{(2)}{}_2O^{(4)}{}_2$ octahedral units. The difference between the T structure and the Or structure is that the distortion is less with a square planar $O_4$ base thereby forming a $Fe^{(2)}O^{(3)}{}_2O^{(4)}{}_4$ octahedral unit. Therefore in the Or structure, one expects two types of formations alternately placed: 1. $Fe^{(1)}O^{(1)}O^{(2)}{}_4$ dimers and 2. $Fe^{(2)}O^{(3)}{}_2O^{(4)}{}_4$ octahedra.

**Table 1:** Reliability parameters from Reitveld refinement of RT-XRD of $SrFe_{1-x}V_xO_{3-\delta}$ ($0 \leq x \leq 3$)

|  | S0 | S1 | S2 | S3 |
|---|---|---|---|---|
| $R_P$ (%) | 2.08 | 2.31 | 2.44 | 2.53 |
| $\chi^2$ (%) | 2.13 | 2.76 | 2.08 | 2.09 |
| Phase (%) | 99.45 (T) | 87.80 (T) | 88.44 (T) | 87.77 (T) |
|  | 0.55 (Or) | 12.20 (C) | 11.56 (C) | 12.23 (C) |
| Vol (Å³) | 922.58 (T) | 927.34 (T) | 928.36 (T) | 928.32 (T) |
|  | 460.74 (Or) | 58.09 (C) | 58.05 (C) | 58.15 (C) |

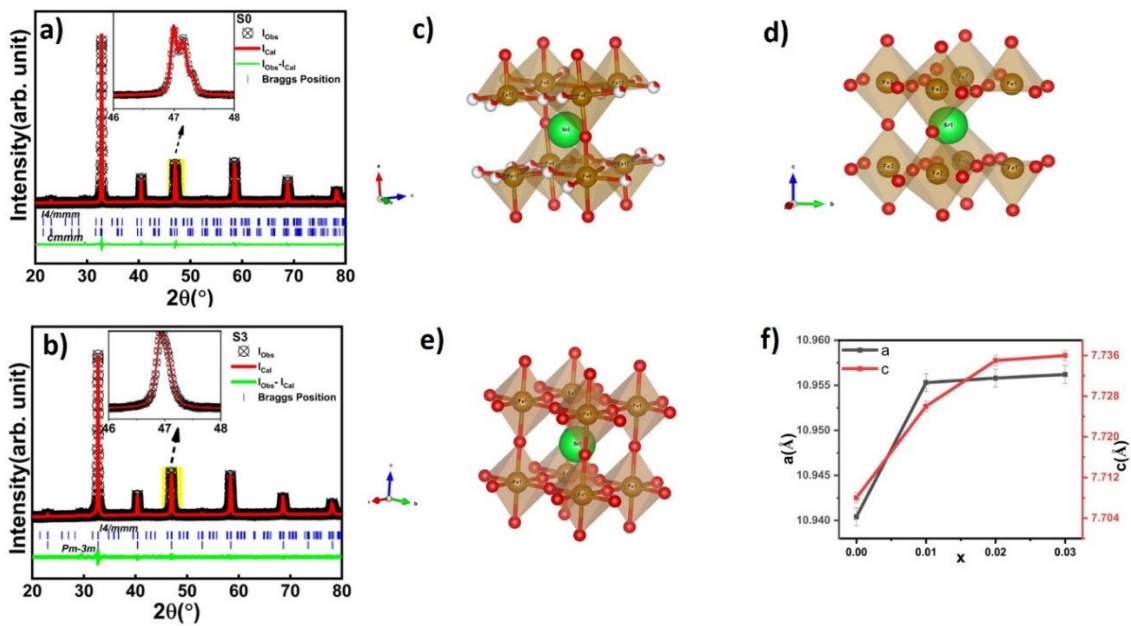

**Fig 2:** Reitveld refinement of a) S0 with *I4/mmm* and *Cmmm* b) S3 with I4/mmm and *Pm-3m* and Vesta generated crystal structure of c) *Cmmm* d) *I4/mmm* and e) *Cmmm* space group and f) variation of lattice parameter obtained from reitveld refinement

Analysis of the major T structure:

There are seven different Fe-O bond lengths as discussed above in the T structure: $Fe^{(1)}$-$O^{(1)}$, $Fe^{(1)}$-$O^{(2)}$, $Fe^{(2)}$-$O^{(3)}$, $Fe^{(2)}$-$O^{(2)}$, $Fe^{(2)}$-$O^{(4)}$, $Fe^{(3)}$-$O^{(4)}$, and $Fe^{(3)}$-$O^{(5)}$. The similarity between the dimers and the octahedra is in the $O_5$ pyramidal structures. To understand the phase transition from a C→T structure one needs to understand the changes manifested in the volumes of these pyramidal structures. Note that in the cubic structure, only one type of pyramid is possible. The volume of these pyramids is ~ 6.61436 Å$^3$. For the tetragonal structure, there are three types of pyramids: one $O^{(1)}O^{(2)}_4$ (dimer unit), b) two $O^{(3)}O^{(2)}_2O^{(4)}_2$ (kite unit), and c) one $O^{(5)}O^{(4)}_4$

(symmetric octahedra units). By calculating the area of these pyramids one observes that the volume reduces for each of them as compared to the C phase: $O^{(1)}O^{(2)}_4 \sim 4.8367 \text{Å}^3$, $O^{(3)}O^{(2)}_2O^{(4)}_2 \sim 5.15673 \text{Å}^3$, and $O^{(5)}O^{(4)}_4 \sim 4.7068 \text{Å}^3$. Hence there is evidence of a strong reduction of pyramidal volume with a cubic to T transition. The ionic radii of the ideal $Fe^{4+}(VI)$ is $\sim 0.725 A$, while that of $Fe^{3+}(VI)$ is $\sim 0.69 A$ for LS and $0.78 A$ for HS. Hence a substitution of $Fe^{4+}$ by HS $Fe^{3+}$ will lead to volume expansion while by LS $Fe^{3+}$ will lead to volume compression. Thus, from these observations one may infer that a C→T phase transition is associated with a oxygen loss due to substitution of $Fe^{4+}$ by LS $Fe^{3+}$.

The V doped samples had a mixture of T (major) and C (minor) phases. By estimating the volume of the pyramids in both phases two observations surfaced. In C phase, the volume increased nominally with increase of doping content but was considerably more than the undoped one. This implies that the presence of V in the C lattice initiates a volume expansion which requires a larger ionic radius than $Fe^{4+}$. Note that the ionic radii of V in different valence state and coordination are as follows: $V^{4+}(VI)$ (0.72A), $V^{5+}(VI)$ (0.68A), $V^{4+}(V)$ (0.67A), $V^{5+}(V)$ (0.60A). Hence apart from $V^{4+}(VI)$, which is comparable to $Fe^{4+}(VI)$, all are smaller. Hence, V alone cannot be responsible for lattice volume expansion. Thereby $Fe^{4+}$ needs to convert HS $Fe^{3+}(VI)$ to ensure a volume expansion. Such is also the observation in the T phase as well. With V incorporation the lattice expands thereby hinting at a $Fe^{4+} \rightarrow$ HS $Fe^{3+}$ conversion. This phenomenon is likely due to a decrease in crystal field splitting, which can result from both an increase in the metal-ligand bond distance and a reduction in the oxidation state of Fe. The increase in bond distance can be attributed to the expansion of the octahedral volume, while the lowering of Fe's oxidation state is confirmed by the XPS results, which will be discussed later.

The crystallite size and lattice micro-strain were calculated using the Williamson-Hall equation: $\beta = 0.9\lambda/ (D \cos\theta) + 4\varepsilon \tan\theta$, where $D$ is crystallite size, $\beta$ is peak broadening, $\varepsilon$ is micro-strain. The calculation was made using the position of the highest intense peak at (*110*) for all four samples. The crystallite size was observed to decrease with an increase in strain from the pure to the doped samples [Supplementary Figure S2]. The differences in the ionic radii of the different valence states of Fe and V and corresponding O content can be accounted for such a change of strain. The increase in strain may be the reason behind the incapability of the lattice to maintain the long-range order.

**Raman Spectroscopy:**

Room temperature Raman spectra were recorded in the frequency range of 100-800 cm$^{-1}$ using a 633 nm laser. For the S0 sample, distinct Raman peaks were observed at 317.9 cm$^{-1}$, 412.3 cm$^{-1}$, and a broad peak between 550-650 cm$^{-1}$ [10]. In the doped samples, these modes were significantly suppressed, suggesting a loss of tetragonality. To understand how these vibrational modes are influenced by doping, it is essential to first determine the origin of the vibrations in the undoped sample. However, the literature on phonon vibrations in SrFeO$_{3-\delta}$ is limited. To address this, zone center calculations were performed using Density Functional Perturbation Theory (DFPT) for the T and Or phases, considering both antiferromagnetic (AFM) and ferromagnetic (FM) configurations. The energy calculations revealed that the magnetic ground state of T-SFO is A-type AFM with a minimum energy of -6.477 eV per atom, while for Or-SFO, the ground state is G-type AFM with a minimum energy of -6.513 eV per atom. The vibrational mode corresponding to the most intense peak at 317.9 cm$^{-1}$ is identified as B$_{1g}$ (T-SFO), associated with the bending motion of oxygen atoms in FeO$_5$ within the ab plane. The next peak at 412.3 cm$^{-1}$ is assigned to the E$_g$ (T-SFO) mode, corresponding to the tilting motion of kite-shaped FeO$_6$ octahedra along the c-axis and intra-polyhedral out-of-phase motion within the regular FeO$_6$ of the T phase. The broad peak above 550 cm$^{-1}$ was deconvoluted into two components: one at 585 cm$^{-1}$ corresponding to the T phase, and another at 625 cm$^{-1}$ corresponding to the Or phase. These modes are attributed to the E$_g$ stretching vibrations in the O-Fe-O bonds of FeO$_5$ and FeO$_6$, and the A$_g$ in-plane bending motion of FeO$_5$ for the T and Or phases, respectively. From this analysis, it is evident that the vibrational modes in the mid-range frequency are primarily influenced by the bending and stretching of FeO$_5$, as well as the tilting and stretching of FeO$_6$. Upon substituting Fe with V, oxygen vacancies are reduced due to the V$^{5+}$ state as compared to Fe$^{3+}$. The Fe$^{3+}$ ion has five 3d electrons while V$^{5+}$ has none. This is a strong reason for reducing the polarization of the V-O bond compared to the Fe-O. This substitution can cause FeO$_5$ in the T-SFO to transition into FeO$_6$, potentially altering the shape of the kite-shaped octahedra. These cumulative effects induce structural symmetry, leading to a notable reduction in the intensity of the Raman peaks, which aligns with our conclusions from the XRD data.

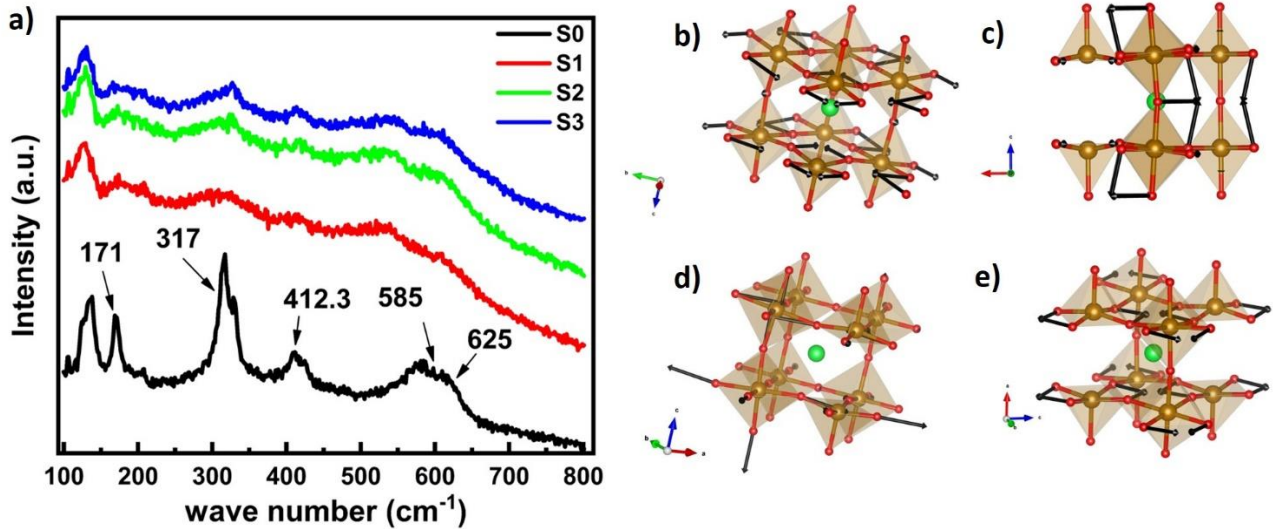

**Fig 3:** a) Raman spectra of all the samples, type of phonon vibration for Raman peak at b) $B_{1g}$ (317 cm$^{-1}$) c) $B_{1g}$ (412 cm$^{-1}$) d) $E_g$(585 cm$^{-1}$) for T-phase e) $A_g$(625 cm$^{-1}$) for Or-phase

## XPS:

XPS survey scans of all the V-doped SFO samples were obtained and the C1s peak was observed at 284.8eV for all the samples. Hence, the data need not be calibrated separately. Detailed scans were performed for Sr 3d, Fe 2p, O 1s and V 2p core level spectra of SFVO series for compositional analysis. A Tougaard background was subtracted from each spectrum. All the high-resolution peaks were deconvoluted to ensure splitting in core levels. In the Fe 2p spectra, two Fe 2p3/2 and 2p1/2 peaks were observed in each spectrum with a spin-orbit coupled separation of $\Delta_{SO}$=13.3 eV [11], [12] . Each feature was deconvoluted into two peaks suggesting a mixed valence state of $Fe^{+4}$ and $Fe^{+3}$. The area ratio of the features $Fe^{4+}(2p_{3/2})$: $Fe^{4+}(2p_{1/2}) = Fe^{+3}(2p_{3/2})$: $Fe^{+3}(2p_{1/2})$ was maintained at 2:1 to account spin multiplicity i.e. 2j+1 of each state. The binding energies of the fitted peaks at ~710.15 eV and ~723.4 eV can be assigned to $Fe^{3+}2p_{3/2}$ and $Fe^{3+}2p_{1/2}$, while, at ~712.4 eV and ~725.7 eV can be the $Fe^{4+}2p_{3/2}$ and $Fe^{4+}2p_{1/2}$. A constant increase in the concentration of $Fe^{3+}$ oxidation state was observed with V-doping [Table:2]. Similarly, for V, the

XPS spectra were fitted with $V^{4+}3p_{3/2}$, $V^{4+}3p_{1/2}$, $V^{5+}3p_{3/2}$, and $V^{5+}3p_{1/2}$ peaks having $\Delta_{SO}=7.3$ eV [13]. With an increase in doping concentration, the ratio of $V^{+5}/V^{+4}$ content increased.

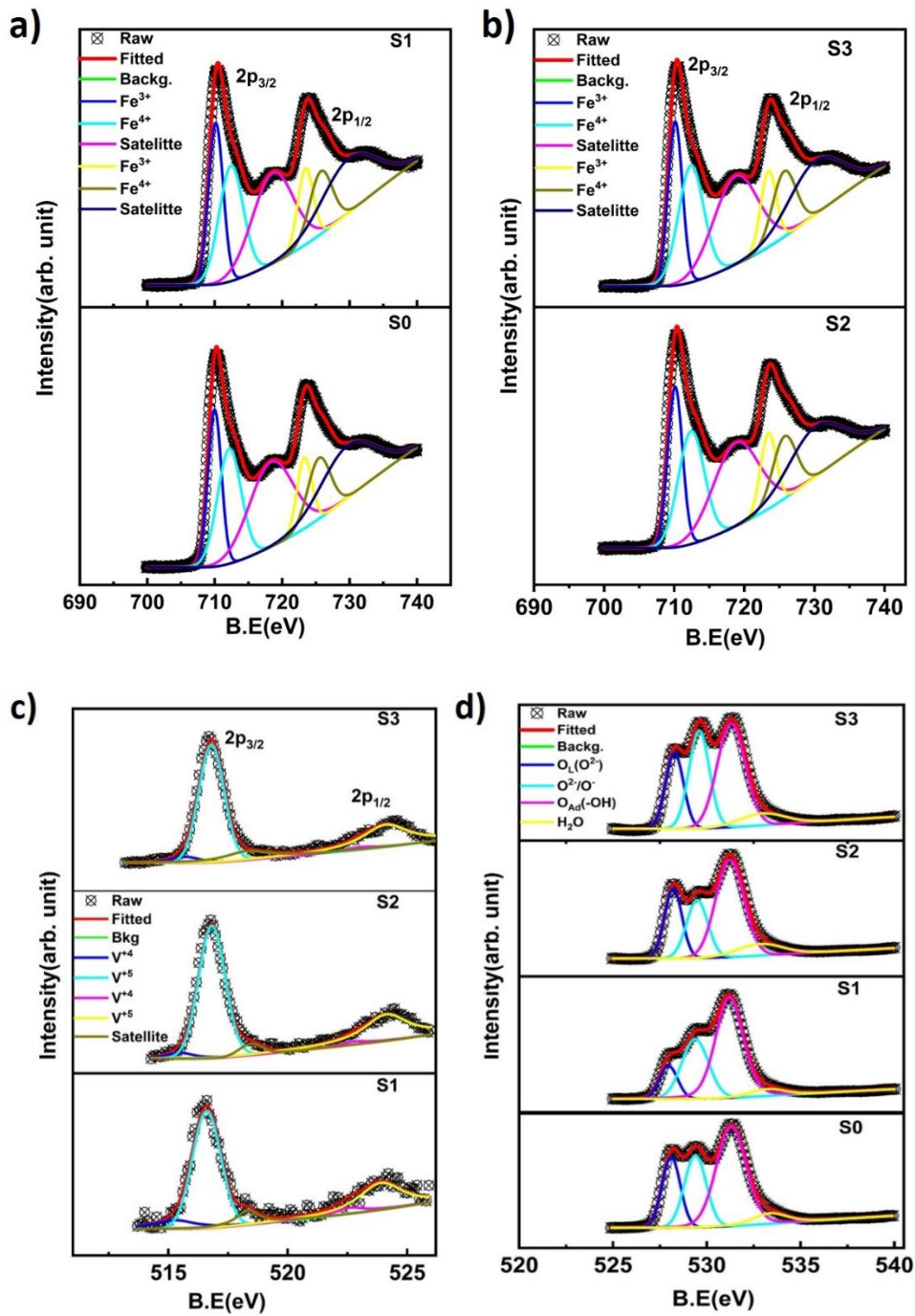

**Fig 4:** Fe 2p deconvoluted XPS spectra for a) S0 & S1 b) S2 & S3, c) V 2p deconvoluted XPS spectra for S1, S2 & S3 d) O 1s deconvoluted XPS spectra for all the samples

The O1s features were deconvoluted into four peaks: lattice oxygen ($O_L$) at 528.1 eV, highly oxidative oxygen ($O^{2-}/O^-$) at 529.4 eV, surface-adsorbed species (-OH/$O_2$) at 531.4 eV, and surface-adsorbed $H_2O$ at 533.4 eV [14]. However, this analysis provided limited insight and was deemed inconclusive. Nevertheless, estimating changes in lattice oxygen is crucial, as the Fe substitution by V affects both elements' valence states. It was observed that $Fe^{3+}$ content increased alongside $V^{5+}$ concentration. A mathematical estimation of the probable cationic valence states was performed to assess oxygen vacancies ($V_O$) in the lattice, revealing a slight reduction in $V_O$. Minor modifications in the oxygen lattice can also influence the Sr lattice due to the large size of Sr ions. Although the valence state of Sr rarely changes, its large size may cause segregation, potentially pushing Sr ions to the surface.

The Sr3d XPS spectra represent predominantly a $Sr^{2+}$ valence state. The spectra could be deconvoluted into two $Sr^{2+}3d_{5/2}$ and $Sr^{2+}3d_{3/2}$ features belonging to the lattice and surface contributions. The lattice contributions are due to the Sr in the perovskite structure while Sr in the surface contributions come from SrO/Sr(OH)$_2$/SrCO$_3$ formation at the surface. The $Sr^{2+}3d_{5/2}$ and $Sr^{2+}3d_{3/2}$ of each contribution had a $\Delta_{SO}$=1.8 eV with an area ratio of 3:2. The $Sr^{2+}3d_{5/2}$ and $Sr^{2+}3d_{3/2}$ features of the lattice were centered at 131.43 eV and 133.23 eV respectively, while for the surface, these were at 133.20 eV and 135.03 eV respectively. The ratio of $Sr_{surface}$/$Sr_{lattice}$ increases with the introduction of V. This may be a consequence of the changes in the O-lattice near the dopant sites.

**Table 2:** Percentage of Fe, V and O species calculated from XPS

| Sample | $Fe^{3+}$ | $Fe^{4+}$ | $Fe^{3+}/Fe^{4+}$ | $V^{+4}$ | $V^{+5}$ | $O_{Lattice}$ | $O_2^-/O^-$ |
|---|---|---|---|---|---|---|---|
| S0 | 0.43 | 0.57 | 0.75 | ----- | ----- | 0.21 | 0.26 |
| S1 | 0.45 | 0.55 | 0.82 | 0.14 | 0.86 | 0.11 | 0.31 |
| S2 | 0.45 | 0.55 | 0.82 | 0.06 | 0.93 | 0.23 | 0.24 |
| S3 | 0.47 | 0.52 | 0.90 | 0.05 | 0.95 | 0.20 | 0.30 |

## **XANES:**

Fe-k edge:

X-ray absorption feature around any edge for most 3d elements surrounded by a ligand, in this case oxygen, are dictated by an overlap of 3d-4p orbitals of the metal as a result of the hybridization of the O-2p orbital. The main edge is due to a dipole ($\Delta l=\pm 1$) allowed 1s-4p transition while the pre-edge is due to quadruple ($\Delta l=\pm 2$) allowed 1s-3d transition. The pre-edge feature is prominent for a distorted structure with unequal bond lengths thereby leading to a larger crystal field splitting, which further introduces a prominence of other nominal features in between the edge and pre-edge. Hence analysis of near edge absorption spectra of a 3d element gives plenty information about not only the oxidation state but the local geometry of the lattice structure as well.

Fe and V K edge XANES spectra were probed to see the local structure around the B site and its evolution with V substitution. The main Fe-K edge peak [Fig:5 (a)] observed at 7130.70 eV due to 1s-4p transition. To verify the edge positions for different valence states of the Fe ions a few normalized room temperature XANES data of the Fe-K edge were obtained for Fe-foil (for $Fe^0$), $FeCl_2$ (as divalent $Fe^{+2}$), $Fe_2O_3$ (as trivalent $Fe^{+3}$), and $Fe_3O_4$ (as mixed valent $Fe^{+2}/Fe^{+3}$) as reference. The Fe-foil sample revealed an edge at 7112 eV, $FeCl_2$ at 7123.64 eV, $Fe_2O_3$ at 7127.726 eV, and $Fe_3O_4$ at 7127.132 eV. Similar Fe-K edge data was also obtained for all the samples [Fig:5 (a)]. Note that the binding energy of the samples has not been affected by the substitution revealing a steady valence state of the Fe ion. The edge was observed to be around 7127-7128 eV which is at a higher energy value than both $Fe_2O_3$ and $Fe_3O_4$. Therefore, the valence state of Fe in these samples is probably higher than 2+ and 3+.

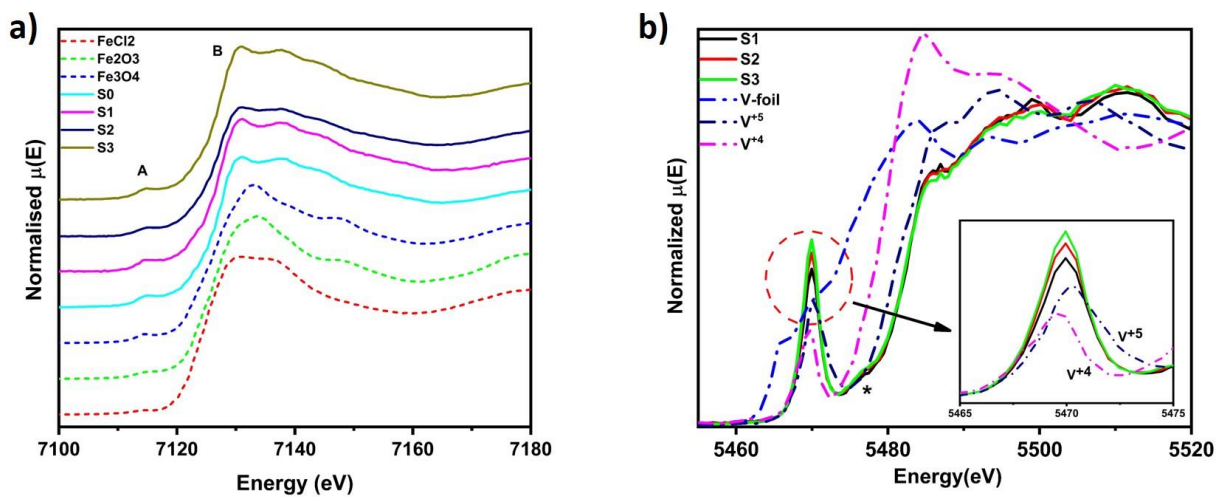

**Fig 5:** Normalized XANES spectra of $SrFe_{1-x}V_xO_{3-\delta}$ obtained at a) Fe-k edge and b) V-k edge

Pronounced pre-edge features [feature B in Fig:5 (a)] are observed at 7115 eV for all doped and undoped samples, thereby revealing the presence of distortion around the Fe centers. These features, as already been described, are due to quadrupole-allowed 1s-3d transition of Fe orbitals. No changes in the intensity or position or shape were observed with variation in x. $Fe^{2+}$ is known not to have a prominent pre-edge peak. However, both $Fe^{+3}$ and $Fe^{+4}$ are known to have sharp pre-edge features. The invariance of the pre-edge feature is also a confirmation of the mixed $Fe^{+3}$ and $Fe^{+4}$ valence state which was also confirmed from the XPS results.

V K-edge:

To further validate the oxidation state of V, the K-edge XANES spectra were collected at room temperature of the doped samples. Reference samples of V-foil, $VOSO_4$ ($V^{+4}$), and $V_2O_5$ ($V^{+5}$) were also collected under a similar environment [Fig:5 (b)]. A strong absorption edge was observed in the energy range 5480-5488 eV due to dipole-allowed 1s-4p transition in vanadium [15]. The position of this feature is located at a higher energy above the $V^{+5}$ feature of $V_2O_5$. This somewhat ensures a predominant presence of V in the $V^{+5}$ state. Note that the pre-edge peak is at

5469.94 eV for all the samples due to a quadrupole V 1s-3d transition. As discussed above the presence of a pre-edge feature generally relates to distortion at the V center with neighboring O. However, the $V_2O_5$ absorption spectra also have a sharp pre-edge feature similar to S1, S2, and S3. Such a pre-edge feature has been reported [15] to be much less and at a lesser energy in the case of $VO_2$ ($V^{+4}$). Hence, from these facts, one can justify the increasing presence of $V^{+5}$ with increasing x. This is in good agreement with the V-2p XPS study.

**Magnetism study:**

The M-H hysteresis loops for SFVO samples were measured at 10 K and 300 K in a magnetic field of ±30 kOe, as shown in Fig. 8. All loops display an unsaturated behavior, indicative of antiferromagnetic (AFM) ordering. A narrow opening of the loop is observed in the doped samples at both temperatures, a feature absent in the undoped sample (S0). The loop width consistently increases with doping of the non-magnetic $V^{5+}/V^{4+}$ ions, as illustrated in the inset of Fig. 8. This emergence of an unsaturated hysteresis loop suggests the presence of ferromagnetic (FM) ordering within the sample. The resultant magnetic ordering of the doped samples is FM superimposed on dominant AFM ordering and the FM part intensifies with percentage of V at low temperature and a mixed of FM and PM at RT. Notably, the loop broadening persists even at room temperature, with both coercive field and remnant magnetization increasing with doping, as seen in Fig. 9.

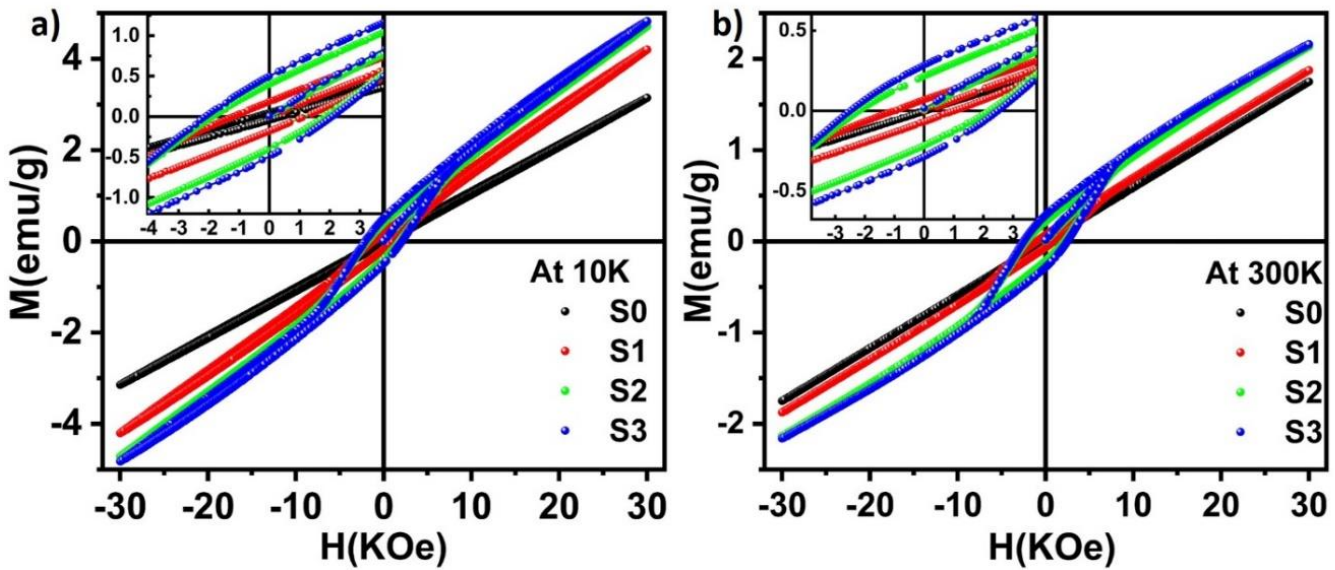

**Fig 6:** Field dependent magnetization of SrFe$_{1-x}$V$_x$O$_{3-\delta}$ obtained at a) 10K and b) 300K

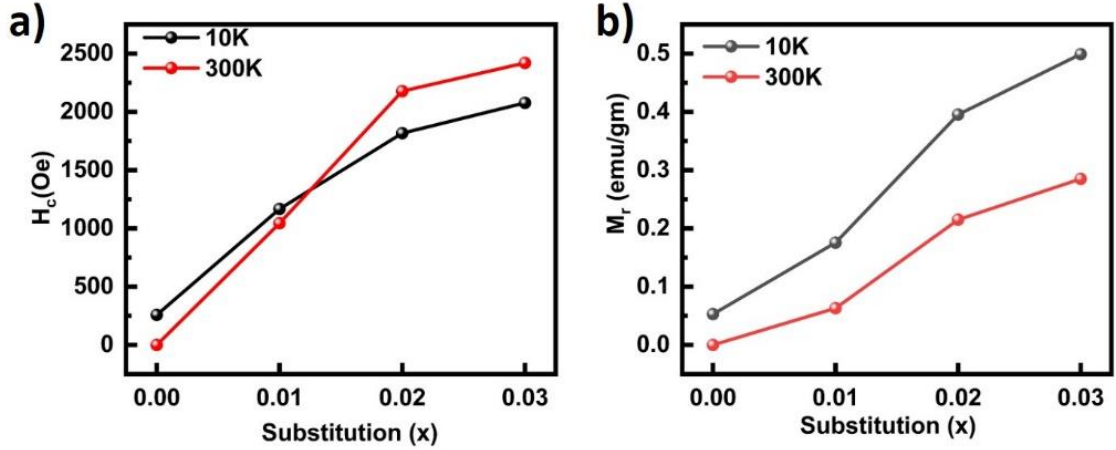

**Fig 7:** Variation of a) Coercive field and b) Remnant magnetization at 10K and room temperature with doping concentration

To understand the effects of V doping on the SFO lattice, probable exchange interactions were examined. From octahedral volume expansion we concluded the presence high spin state of Fe$^{3+}$ (t$_{2g}^3$e$_g^2$) whose concentration increases as V$^{5+}$ (t$_{2g}^0$e$_g^0$) increases. The possible exchange interactions for AFM ordering is likely come from super-exchange (SE) interactions such as Fe$^{3+}$-O-Fe$^{3+}$ and Fe$^{4+}$-O-Fe$^{4+}$. In contrast, FM contributions may stem from Fe$^{3+}$-O-Fe$^{4+}$ double exchange (DE) interactions. XPS studies reveal an increase in the Fe$^{3+}$ fraction from 0.43 to 0.47, while the Fe$^{4+}$ fraction decreases from 0.57 to 0.52. This in turn enhances Fe$^{3+}$-O-Fe$^{4+}$ DE interactions responsible for FM behavior. The AFM contribution correspondingly decreases due to a reduction in Fe$^{4+}$-O-Fe$^{4+}$ SE interactions. Additionally, DE interactions between V$^{5+}$-O-Fe$^{3+}$ could further support parallel spin alignment, reinforcing FM ordering in the sample.

The presence of competing magnetic orderings provides a potential basis for the emergence of spin frustration within the lattice structure [16]. To further investigate this, temperature-dependent DC and AC measurements were conducted under Zero Field Cooling (ZFC) and Field Cooling (FC) conditions to analyze the effect of V doping. Previous studies on pure SFO have identified several features in the 70 K to 230 K range within ZFC and FC data, which have been predominantly associated with various AFM orderings. A commonly observed feature is a cusp or sharp maximum around 70 K, indicative of AFM ordering in the T- $Sr_8Fe_8O_{23}$ phase, largely comprised of C-SFO structures lacking one oxygen atom per eight-unit cells. This sharp feature around 70 K is commonly attributed to magnetically disordered $Fe^{4+}$ sites. Divergence in FC and ZFC data below $T_N$ is a typical characteristic of the T-SFO phase, which becomes more pronounced with increased oxygen loss [17]. Another, frequently reported feature appears around 140 K, associated with a helical spin AFM order [3]. This feature is typically observed in C-SFO. A structural transition is thus inferred to influence the nature of magnetic ordering. For Or-$Sr_4Fe_4O_{11}$ lattices with increased oxygen content, an irreversible feature below 240 K is observed [18].This behavior is indicative of antiferromagnetic ordering of $Fe^{3+}$ sites, representing a well-documented magnetic phase transition correlated with the Or-phase [19].

The ZFC/FC measurements of the samples under DC magnetic fields of 200 Oe across a temperature range of 5 K to 300 K is shown in Fig. 7 (a)–(d). For the undoped sample (S0), ZFC and FC curves start to diverge below 232 K when the applied field in 200Oe (indicated in the inset of Fig. 7 (a) as Tirr). Tirr shifts to lower temperatures as the applied field increases to 500Oe and 1KOe. Beyond Tirr, both ZFC and FC magnetizations reach a maximum around 70 K. For the doped samples, peaks are observed around 55 K, and no irreversibility like that in S0 is observed beyond the Néel temperature ($T_N$). As discussed previously in S0 the maxima at 70K is due to T-phase and the irreversibility around 232K is due to the Or-phase. The shift of $T_N$ value to a lower temperature ~ 55K for all the doped samples indicates weakening of AFM ordering from S0. The difference in magnetization between ZFC and FC, $\Delta M_{ZFC-FC}$ below $T_N$ increases progressively from sample S1 to S3 which clearly suggest a magnetic frustration in the lattice. Variation of $\Delta M_{ZFC-FC}$ with temperature [Fig: S3 supplementary] shows that there is no irreversibility of the doped sample beyond $T_N$. The interaction between FM (interlayer) and AFM (intralayer) orders leads to the observed differences in ZFC and FC curves, i.e., $\Delta M_{ZFC-FC}$ which are comparable to the behavior seen in the Or- $Sr_4Fe_4O_{11}$ phase [19].

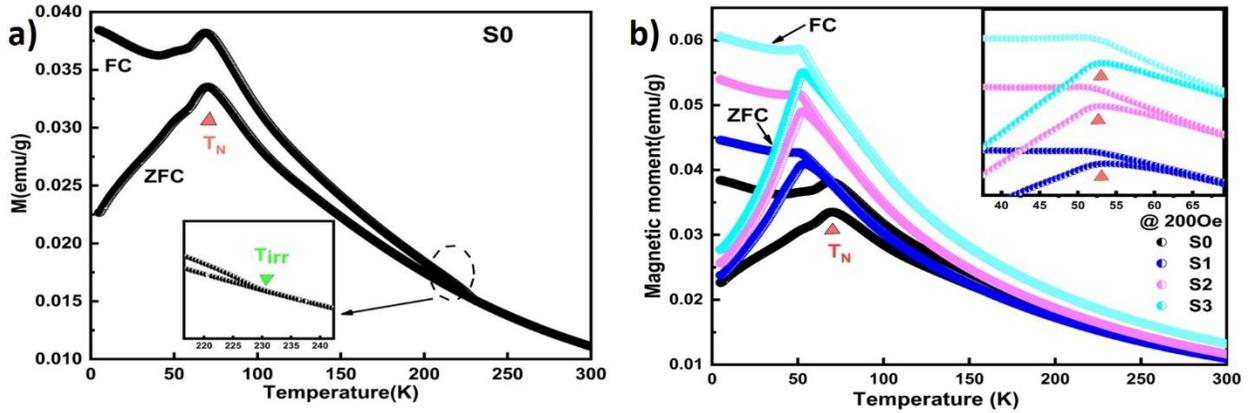

**Fig 8:** FC-ZFC magnetization measured a) for S0 sample at at 200 Oe b) for all the samples at 200 Oe

Further, the spin glass (SG) state was investigated with AC susceptibility measurement at 200Oe [Fig: 10]. It was observed that for the pure SFO the feature at 70K does not shift with frequency. This indicates a non-SG nature of the sample. However, with V doping with increasing frequency shows a nominal shift towards higher temperature. doping [Fig:8]. Hence V doping introduces FM as well as a weak SG nature at the compromise of the AFM component. This is possible when the long-range AFM order of the Fe-Fe is affected by the presence of V in the lattice. The interaction of $Fe^{3+}$ $e_g$ electrons with neighboring vacant d-orbital of $V^{5+}$ ions can generate an uncertainty in spin orientation which may result in SG system. This interaction in the pure system. Note that from the XRD studies a decrease of the crystallite size was observed with V. The decrease of the crystalline size is consequence of the increase in the lattice strain. the increase in the lattice strain is an indication of the presence of V and a consequent change in the environment of the B site. This is associated with a phase change along with the compensation of the O-loss which maybe contributing to the changes in magnetization especially the glassy nature of the materials.

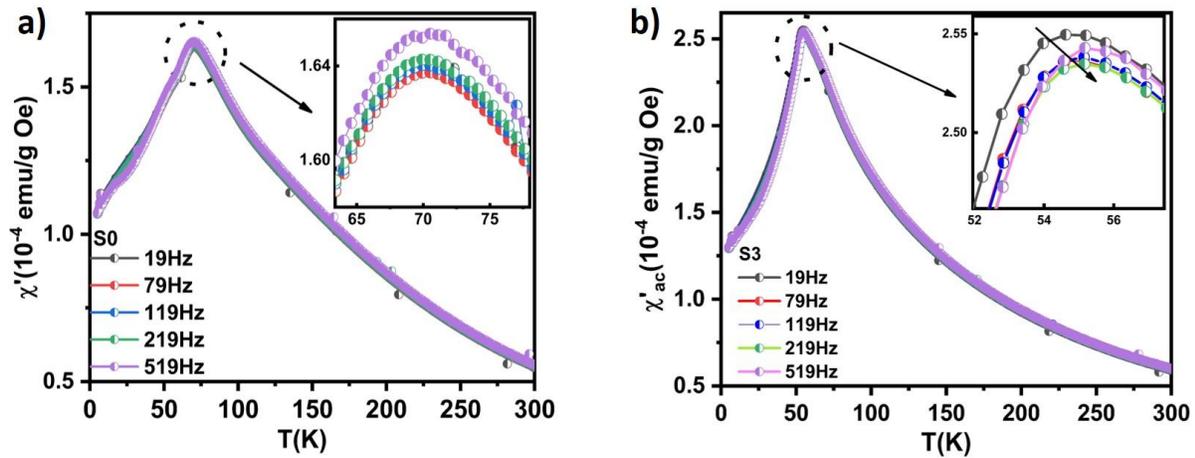

**Fig 9:** AC susceptibility measurement for a) S0 and b) S3 sample

## Conclusion:

Vanadium has been successfully substituted at the iron site in $SrFe_{(1-x)}V_xO_3$ (x=0,0.01,0.02,0.03) by modified Pechini sol-gel method. X-ray diffraction (XRD) and Raman spectroscopy reveal a structural transition towards greater symmetry. The S0 sample shows a major tetragonal phase with a minor orthorhombic structure. Where as the doped sample shows cubic nature along with the T-phase. Reitveld refinement shows a increase in octahedral volume, which suggest a change in spin state of $Fe^{3+}$ to LS to HS. XPS and XANES study indicates a mixed valence state of $Fe^{3+}/Fe^{4+}$ and $V^{4+}/V^{5+}$. The $Fe^{3+}$ and $V^{5+}$ concentration increases with doping and which effects the oxygen vacancy in the lattice. For x=0.01,0.02 and 0.03 the M-H data shows a weak ferromagnetism in the dominant AFM state at LT and RT both. The intrinsically non-magnetic $V^{5+}$ ($d^0$) induces ferromagnetism by hybridization with O-2p state. This enhancement in FM is been linked with exchange interaction with DE $Fe^{3+}$-O- $Fe^{4+}$ and SE $Fe^{3+}$-O- $V^{5+}$ which facilitates parallel alignment of spins. The $T_N$ value from 70K to 55k in the doped sample shows the net weakening in AFM state along with increased magnetic frustration in the lattice. The increase in difference in magnetization of FC/ZFC from S1 to S3 indicate the long-range magnetic order might have been destroyed by V in Fe-chain. The slight shift in the susceptibility peak value with frequency confirms the presence of spin glass in the lattice.